\documentclass[twocolumn,prl,floatfix]{revtex4}
\usepackage{color}
\usepackage{epsfig}
\usepackage{graphicx}
\usepackage{amsmath}
\begin{document}
\title{Nonlinear Two-Dimensional Green's Function in Smectics}
\author{ E A. Brener$^{1}$, V. I. Marchenko$^{2,3}$  and D. Pilipenko$^{1}$,\\
$^{1}$ Institut f\"ur Festk\"orperforschung, Forschungszentrum
J\"ulich, D-52425 J\"ulich, Germany\\
$^{2}$ Kapitza Institute for Physical Problems, Russian Academy of Sciences, ul. Kosygina 2, Moscow, 119334 Russia \\
$^{3}$ Moscow Institute of Physics and Technology, Institutskii per. 9, Dolgoprudnyi, Moscow region, 141700 Russia 
}
\date{\today}
\begin{abstract}
The problem of the strain of smectics subjected to a force distributed over a line in the basal plane has been 
solved. 
\end{abstract}
\maketitle
The asymptotic expressions for strains around isolated defects in smectics at long distances are characterized by the exponent $\alpha$  \cite{BM07}.
If ${\alpha<0}$,  the linear theory is applicable. If ${\alpha=0},$ (edge dislocation \cite{BM,IL}, GreenÕs function  \cite{BM09}), the linear solution is valid at 
small action amplitudes, whereas nonlinear effects 
become important at larger amplitudes. If ${\alpha>0},$ nonlinear effects should be taken into account even for 
extremely weak actions.  In this paper, we report the solution of the nonlinear problem of the two-dimensional GreenÕs function ($\alpha = 1/2$)  \cite{BM07}. 
Let us consider a smectic sample with the thickness $L$ that is sandwiched between solid undeformable walls 
parallel to smectic layers (see figure). A force uniformly distributed along the y axis (normal to the figure plane) with the linear density $F$
 is applied at the center ($x=z =0$) of the smectic sample. 
The energy of the small strains of the smectic sample is given by the expression (see Eq. (44.13) in \cite{Landau})
\begin{equation}\label{E} 
E=\frac{A}{2}\int\left\{\left(\partial_zu -\frac{(\partial_x u)^2}{2}\right)^2+\lambda^2(\partial^2_x u)^2\right\}dV,
\end{equation}
where $u$ is the displacement of the layers along the smectic axis $z$, $A$  is the elastic modulus, and $\lambda$ is the 
microscopic length parameter. In our problem, the maximum displacement $u_0$ is reached in the force 
application line. 
In terms of the new function $f = u/u_0$ and new coordinates $\tilde{z} = z/L$ and $\tilde{x}=x/ L$, where 
$\varepsilon=u_0/L$, Eq. ~(\ref{E}) is represented in the form
\begin{equation}\label{E1} 
E=2^{-1}AL^2\varepsilon^{5/2}\int\left(\sigma^2
+\beta(f'')^2\right)d\tilde{z}d\tilde{x},
\end{equation}
where ${\beta=(\lambda/\varepsilon L)^2},$~ ${\sigma=\dot{f}-(f')^2/2},$ and the dot and 
prime mean differentiations with respect to  $\tilde{z}$ and  $\tilde{x}$
respectively. Thus, it is necessary to determine the 
function $f$ that provides the minimum of energy (\ref{E1}), is 
equal to unity at ${\tilde{x}=\tilde{z}=0}$, and is equal to zero at the 
edges of the smectic layer ${\tilde{z}=\pm{1/2}}.$ The force is given 
by the expression ${F=F_+-F_-}$ , where 
\begin{equation}\label{F}
F_\pm=\mp\int\sigma_{zz}dS=\mp AL\varepsilon^{3/2}\int\sigma d\tilde{x}.
\end{equation}
\begin{figure}
\epsfig{file=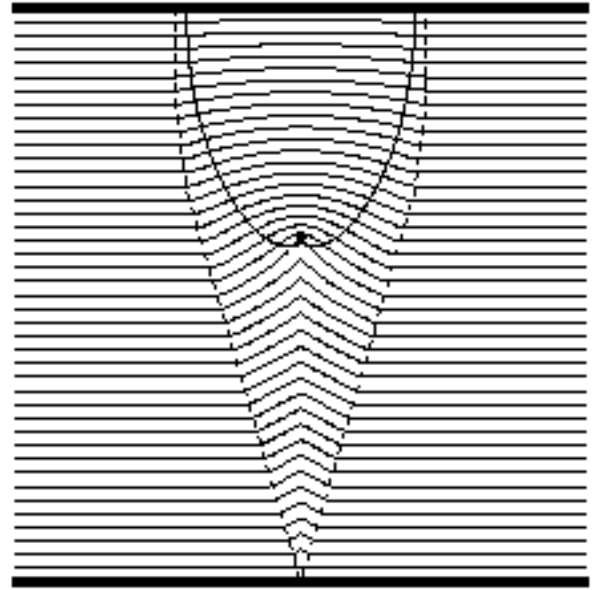, width=10cm,angle=-0}
\caption{Strain of the smectic layer subjected to the linearly distributed force(${\varepsilon=0.1}$)applied at the thick point. The dashed 
line is the boundary of the region, where a noticeable strain 
appears. The thin line is the boundary of the compression 
region($\sigma<0$).}
\end{figure}

In the macroscopic problem, curvature  ${\propto\beta}$ can be 
neglected even for the case of a negligibly small force ${F\sim\lambda A\sqrt{\lambda/L}}$, where the amplitude $u_0$ is larger than the 
distance between smectic layers ${\sim\lambda\ll L}$. It is interesting that integrals (\ref{E1}) and (\ref{F}) can be calculated in this 
case even without the complete solution, because the 
strain field is divided into two regions. In the first 
region, where the material is compressed, the problem 
can be solved analytically. In the second region, owing 
to the Helfrish instability (see the problem in Section 44 in \cite{Landau}), rotary states almost without stresses appear 
instead of tension. Here, noticeable stresses exist only 
inside microscopically thin twin boundaries  \cite{M}. 

The figure shows the strain pattern in the smectic 
layer obtained by numerically solving the problem. In 
the compression region, the function $f$ is a quadratic 
function of the variable $\tilde{x}$: ${f(\tilde{x},\tilde{z})=
f_0(\tilde{z})+f_2(\tilde{z})\tilde{x}^2.}$
It is easy to verify that the equilibrium equation   ${\dot{\sigma}-(f'\sigma)'=0}$ corresponding to the extremum of energy 
(\ref{E1}) has such an exact solution satisfying the necessary conditions. It can be represented in the parametric 
form 
\begin{equation}\label{f}
f_0=1-\frac{2\varphi}{\pi}, f_2=-\frac{\pi\cos\varphi}{2\sin^3\varphi},
\tilde{z}=\frac{\varphi-\sin\varphi\cos\varphi}{\pi}.
\end{equation}

The compression region ${\sigma<0}$  corresponds to the 
range ${-\tilde{x}_m<\tilde{x}<\tilde{x}_m}$, at  ${\tilde{z}>0},$ where  ${\tilde{x}_m=2\sin^2\varphi/\pi}.$
The thin line in the figure is the boundary taking into 
account the displacement of smectic layers, where $\sigma$ is 
zero. Under this line, down to the dashed line, strains 
result in the appearance of two twins \cite{M}  in which 
inhomogeneous strains increase in the direction to the 
force application level. Here, stresses are small and are 
likely caused by the finiteness of the grid used for the 
calculation. 

When the tension of the twin boundaries is disregarded ( at ${\beta=0}$) \cite{M}, ${F_-=0}.$ Then,
${F=F_+=(8/3\pi)AL\varepsilon^{3/2}}.$ Correspondingly, 
\begin{equation}\label{u0}
u_0=\left(\frac{3\pi a}{8}\right)^{2/3}L^{1/3},\, {a=\frac{F}{A}}.
\end{equation}
At ${\tilde{z}\ll1},$ according to (\ref{f}), $u=u_0+\delta u,$ where
\begin{equation}\label{u+}
\delta u=-3\left(\frac{a^2z}{16}\right)^{1/3}
-\frac{x^2}{3z}, \, {\delta u\ll u_0}.
\end{equation}

Let us assume that the displacement near the force 
application point has the form ${\delta u=-z^{1/3}a^{2/3}\psi(v),}$
where ${v=xa^{-1/3}z^{-2/3}}.$ Note that $\delta u$is independent of 
$L$. In this case, the equilibrium equation is the ordinary differential equation 
\begin{equation}\label{psi}\left(\left(2\psi-4v\psi'+
3(\psi')^2\right)\left(3\psi'-2v\right)\right)'=0,
\end{equation}
where the prime means differentiation with respect to 
 $v$, and has a trivial first integral. The integration constant is zero, because the expression in the second 
parentheses in Eq.~(\ref{psi}) is zero for solution (\ref{u+}). The 
expression in the first parentheses is proportional to 
stress $\sigma$. It is convenient to represent the solution of 
the equation $\sigma =0$  in the parametric form 
${\psi=(4+t^3)/2t},$ and  ${v=(1+t^3)/t^2.}$ The intervals ${-1<t<0}$
and ${0<t<2^{1/3}} \rightarrow{z>0}$ correspond to the regions $z<0$ and 
$z >0$ respectively. The integration constant is chosen 
from the condition ${\psi\left(3/2^{2/3}\right)=3/2^{1/3}}$ at  ${t=2^{1/3}}$ (matching with solution (\ref{u+}) on the line ${x_m(z)\simeq3(az^2/4)^{1/3}}$).
Near the ${z=0}$ line (${|v|\rightarrow\infty}$),
$$\delta u=-2\sqrt{a|x|}+\frac{az}{2|x|}.$$

At the  ${x=0}$ line at ${z<0}$ (${|v|\rightarrow0}$)
$$\delta
u=-\frac{3}{2}a^{2/3}|z|^{1/3}-\frac{a^{1/3}|x|}{|z|^{1/3}}.
$$
If ${a\ll\lambda},$ onlinear asymptotic expressions are valid at ${|z|\gg\lambda^3/a^{2}},$ and ${|x|\gg\lambda^2/a},$
and the GreenÕs function of the linear approximation is applicable at smaller 
distances \cite{BM07}. If ${a\gg\lambda},$ small-strain approximation (\ref{E}) is  violated at ${|z|\sim|x|\sim a}$.

Stresses in the problem under consideration exist 
only in the smectic compression region and rotary 
adjustment in a certain bounded region occurs instead 
of tension. Such a character of nonlinear response 
implies that the compression field and, correspondingly, the $u_0$ value (with the change $L\rightarrow2L_+$), as well 
as the asymptotic expression for $\delta u$, remain 
unchanged in the general case, where the force is 
applied not to the center, but at any distance $L_+$ in the 
direction of the force action from the undeformable 
wall. The boundary conditions on the opposite side 
and, generally speaking, beyond the compression 
region affect only the adjustment structure at distances 
of about the sample sizes. 
This work was supported by the German-Israeli 
Foundation and the Russian Foundation for Basic 
Research (project no. 09-02-00483).

\end{document}